\documentstyle[twocolumn,epsf]{jpsj}

\def\sf{\rm}

\def\runtitle{Effects of Nonmagnetic Impurity Doping on Spin Ladder System}
\def\runauthor{Youichirou {\sc Iino} and Masatoshi {\sc Imada}}

\title
{
	Effects of Nonmagnetic Impurity Doping on Spin Ladder System
}

\author
{ 
Youichirou {\sc Iino}
and Masatoshi {\sc Imada}
}

\inst
{
  Institute for Solid State Physics,\\
  University of Tokyo, Roppongi 7-22-1,\\
  Minato-ku, Tokyo 106
}

\recdate
{
}

\abst
{
Effects of nonmagnetic impurity doping on an antiferromagnetic 
spin-$1/2$ Heisenberg ladder system are studied by 
the quantum Monte Carlo method.
A single nonmagnetic impurity induces a localized spin-1/2 moment 
accompanied by {\it static} and enhanced antiferromagnetic 
correlations around it.
Small and finite concentration of impurities induces a remarkable change of
magnetic and thermodynamic properties with gapless excitations.
It also shows rather sharp but continuous crossover 
around the concentration of about 4\%. 
Above the crossover concentration, all the spins are strongly coupled 
participating in the enhanced and rather uniform power-law decay of 
the antiferromagnetic correlation.
Below the crossover, each impurity forms an antiferromagnetic 
cluster only weakly coupled each other.
For random distribution of impurities, large Curie-like susceptibility 
accompanied with small residual entropy is obtained 
at low temperatures in agreement with recent experimental 
observation in  ${\rm Zn}$-doped ${\rm SrCu_{2}O_{3}}$.
Temperature dependence of antiferromagnetic susceptibility shows 
power-law-like but weaker divergence than the single 
chain antiferromagnet in the temperature range studied.
}

\kword
{
spin ladder system, spin gap, ${\rm SrCu_{2}O_{3}}$, 
impurity effect, quantum Monte Carlo method, crossover, randomness
}

\input defphys.inc

\begin{document}
\sloppy
\maketitle

Spin-gapped Mott insulators have recently attracted much attention since the 
discovery of high-temperature superconductors with pseudo-spin-gap phase 
near the Mott insulator.
A typical theoretical model for the spin-gapped Mott insulator is found in the 
antiferromagnetic Heisenberg (AFH) model with 2-leg ladder shape (two coupled 
spin chains)  where a large spin gap is basically due to the dominant singlet 
formation on the rungs \cite{Dagotto,White}
The spin gap is numerically estimated as $\Delta\sim0.5 J$ \cite{White} 
for the uniform exchange coupling $J$.
Experimentally, the ladder model is believed to be relevant for systems such as 
${\rm (VO)_{2}P_{2}O_{7}}$ and ${\rm SrCu_{2}O_{3}}$. 
Recently, effects of ${\rm Zn}$-doping on 
${\rm SrCu_{2}O_{3}}$ (${\rm Sr(Cu_{1-x}Zn_{x})_{2}O_{3}}$ ) 
have been studied where ${\rm Zn^{2+}}$ ion substitutes 
${\rm Cu^{2+}}$ ($S=1/2$) and
plays a role of nonmagnetic impurity.
\cite{Exprm0,Exprm1}
Even small concentration of nonmagnetic impurities causes a drastic 
change of magnetic properties with disappearance of the spin gap. 
At low temperatures below 10K, the ${\rm Zn}$-doping induces anomaly 
of the susceptibility 
which is perhaps due to the antiferromagnetic (AF) transition. 
Above the anomaly of presumable 
Neel temperature, ${\rm Zn}$-doping performed for the concentration 
$0.01\simle x\simle 0.08$ shows the Curie-like 
susceptibility which is consistent with the formation of free localized 
moments proportional to the number of impurities. 
On the other hand, the $T$-linear dependence of the specific heat is 
observed upon doping above $T_{\rm N}$ with the $\gamma$ value per 
spin similar to that of 1D AFH chain (=$2/3J$). 
The low temperature specific heat implies that the residual entropy 
just above $T_{\rm N}$ is inconsistently smaller than that 
is expected from the free localized spins with the Curie law because 
$J\sim 10^{3}$K or larger.
Recently exact diagonalization of clusters and the variational Monte Carlo 
calculations have been performed for the ladder model 
with regularly distributed nonmagnetic impurities.\cite{Motome}  
It suggests quick collapse of bulk gap 
structure in the excitation spectra with sensitive enhancement of 
AF correlations for small impurity concentration consistently 
with experimental indications.
This model has also been studied with the help of mapping to 
the random bond model and the appearance of the Curie-law has been suggested 
in three different regions of temperatures.\cite{SigFuru}

In this paper, to understand these somewhat puzzling experimental 
results with sensitive 
dependence on the impurity concentration,  
depleted ladder system is studied using an unbiased 
numerical method for large system sizes.  We show results of Monte Carlo 
calculations for the ladder model with nonmagnetic impurities in both 
cases of regular and random distributions of impurities.  
In our results, large Curie-like susceptibility and small residual entropy are 
consistently understood within this model, where the overlap of 
AF clusters formed around impurities is substantial.
We also show that rather sharp crossover exists around the 
impurity concentration $x_{\rm c}\sim 0.04$.  
This is due to the fact that the overlap of AF clusters formed 
around the impurities becomes exponentially 
weak below this impurity concentration. 
To understand the low temperature 
properties for $x<x_{\rm c}$, we point out the necessity to take 
account of three dimensional inter-ladder coupling 
to explain the experimental indications even above $T_{\rm N}$.

Our model is 2-leg ladder AFH model with periodic boundary condition 
given by the Hamiltonian, 
$
	H=J\sum_{\langle i,j \rangle}S_{i}\cdot S_{j},
$
where the summation is taken for all pairs of nearest neighbor sites.
To take into account the nonmagnetic impurity effect, 
we deplete the spins in the impurity sites.
For the convenience of notation, we define an index of site as $(x,y)$,
where $x$($y$) is the coordinate in the leg(rung)-direction of the ladder.
For instance, for 64$\times$2-site system, 
$x$ is from $1$ to $64$ and $y$ is from $1$ to $2$.
The site $(x,y)$ belongs to A(B)-sublattice when
$x+y$ is even(odd) number. 

We adopt the quantum Monte Carlo (QMC) method 
with an efficient loop cluster algorithm. \cite{Loop0,Loop1}
This algorithm together with the improved estimators
reduces the statistical errors of the observables and the simulation time.
We treat 64$\times$2-site system with 0, 2, 4, 8, 10 and 16 impurities 
(namely, impurity concentration 
$x$=0, 0.0156, 0.0313, 0.0625, 0.0938, 0.125), 
and 60$\times$2-site system with 10 impurities ($x$=0.0833).
The lowest temperature in our simulation is $1/120$ in the scale of $J$. 
In terms of experiments in the Cu oxide ladders, 
this lowest temperature corresponds to 
the order of 10K which is indeed relevant when one 
wishes to compare with experimental results.
Simulations are performed with $10^{3}$-time sweeps for
equilibration and 5$\times$$10^{4}$-time sweeps for measurements.
The autocorrelation time of the loop cluster algorithm 
is typically a few sweeps.
In case of the random distribution of impurities, we take typically 
average over 72 configurations of samples.

\begin{figure}
\epsfxsize=8.2cm \epsfbox{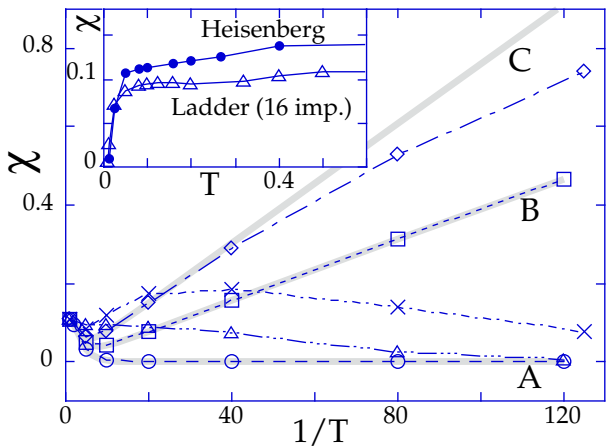}
\caption{$1/T$-dependence of the uniform susceptibility 
for 0($\circ$), 2($\Box$), 4($\Diamond$), 8($\times$), and 16($\triangle$) 
impurities on 64$\times$2-site system. 
Lines A, B and C represent the fitting by the theoretical form. 
Other dashed lines are drawn as a guide to eye.
The inset shows $T$-dependence of the uniform 
susceptibility for 64site AFH chain($\bullet$) 
and ladder system with 16 impurities($\triangle$). }
\label{fig:kai}
\end{figure}

First we investigate effects of nonmagnetic impurities
distributed regularly, where one impurity is located at site (1,1)
and other impurities are put on two legs alternatingly with the same interval. 
In Fig.\ref{fig:kai}, we show the temperature dependence of 
susceptibility 
$\chi=T^{-1}N_{s}^{-1}\sum_{i,j} \langle S^{z}_{i}S^{z}_{j}\rangle$, 
where $N_{s}$ is the number of sites.
This figure shows the qualitative difference between 4- and 8-impurity.
Without an impurity, because of the spin gap $\Delta$, 
susceptibility becomes zero at low temperatures in accordance with 
the theoretical form \cite{Tsune}
$\chi(T)\sim C_{0} T^{-1/2}\exp(-\Delta/T)$ (Line A), 
where $C_{0}$ is a constant.
With 2 and 4 impurities, within the temperature range we studied, 
the susceptibility is well described by 
$\chi(T)\sim C_{0} T^{-1/2}\exp(-\Delta/T) + C_{1}(n)/T$ (Lines B and C), 
where $C_{1}(n)=n/4 N_{s}$ is the Curie constant which is 
proportional to the number of impurities $n$.
The spin gap is estimated as $\Delta=0.497\pm 0.001$ 
in the pure system while the parameter $\Delta$ is estimated as 
$\Delta=0.526\pm 0.001$ and $\Delta=0.553 \pm 0.001$ 
for 2- and 4-impurity cases, respectively.
At $x<x_{\rm c}(\sim 0.04)$,
each impurity generates one spin-1/2 moment 
in the sea of spin-gapped background. 
Because the coupling between these spin moments 
is weak, they behave as nearly free in this temperature range. 
For example, for 2-impurity case at $1/T$=40, the two impurities are 
sufficiently well separated so that we can see effects of 
single-impurity doping around one of it.
Each isolated impurity induces
$S$=1/2 spin. In addition it generates static correlation 
whose saturated value at large $\tau$ for sites $r_{0}$=($1$,2) and 
$r_{j}$=($j$,2) around the impurity (1,1) shows
an exponential dependence on $|r_{0}-r_{j}|$
as $\langle S^{z}_{0}(\tau$=20)$S^{z}_{j}(0)\rangle\propto
(-1)^{|r_{0}-r_{j}|}\cdot\exp(-|r_{0}-r_{j}|/\xi)$
with $\xi$=2.91$\pm$0.15, which shows good agreement 
with the correlation length of pure ladder system, 
$\xi\simeq 3.19$. \cite{Tsune}
Therefore it is concluded that an impurity induces $S$=1/2 spin as well as 
{\it static} and enhanced AF correlation within the range of 
original correlation length of the gapped system.
Similar conclusions have been reached by other theoretical studies.
\cite{Martins,Fukuyama}

In contrast to this low impurity concentration, 
at $x>x_{\rm c}$,
the susceptibility does not show the 
Curie-law behavior in this regular distribution of impurities. 
As is shown in the inset of Fig.\ref{fig:kai}, 
it is rather similar to that in the 1D pure AFH with 
gapless spin excitation. (Note that the actual data 
suffer from finite-size effect at very 
low temperatures with vanishing susceptibility as $T\rightarrow 0$).
Spins localized around impurities 
become strongly coupled above the crossover concentration 
$x_{\rm c} \sim 0.04$ in the temperature range studied.

\begin{figure}
\epsfxsize=8.2cm \epsfbox{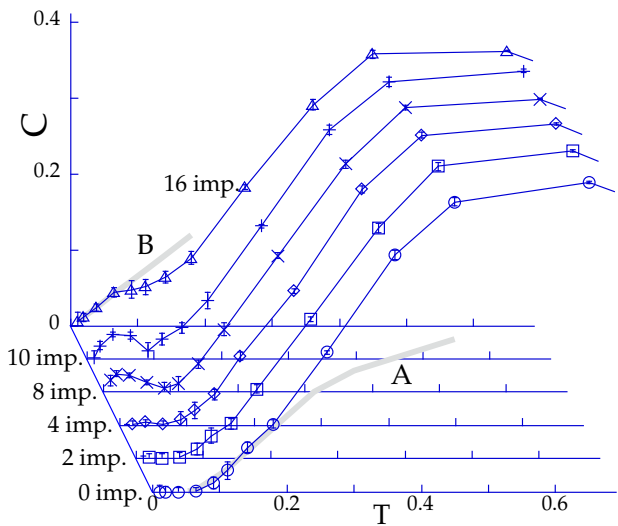}
\caption{$T$-dependence of specific heat per spin 
for 0, 2, 4, 8, 10 and 16 impurities. 
Line A shows the theoretical form for the pure ladder at low $T$.
Line B represents the specific heat per site for 1D AFH chain.
Other lines are drawn as a guide to eye. }
\label{fig:sph}
\end{figure}

In Fig.\ref{fig:sph}, temperature dependence of the specific heat 
per spin is presented.
For the pure ladder system,   
if we assume the momentum dependence of the low energy triplet dispersion
as $\epsilon(k)=\Delta+a|k-\pi|^{2}$, 
with $a$ being a constant,
low temperature ($T\ll\Delta$) form of the specific heat \cite{Tsune} 
is given by 
$
	C(T) = \frac{3}{4}\left(\frac{\Delta}{\pi a}\right)^{1/2}
	\left[ \left(\frac{T}{\Delta}\right)^{-3/2}
	+ \left(\frac{T}{\Delta}\right)^{-1/2}
	+ \left(\frac{T}{\Delta}\right)^{1/2} \right]
	e^{-\Delta/T}
$.
This form fits well to our QMC data for $T\simle 0.2$, 
with the spin gap $\Delta$ estimated as $0.49\pm 0.04$.
At $x<x_{\rm c}(\sim 0.04)$, 
temperature dependence of the specific heat 
looks similar to the pure ladder case.
For these dopings, as is seen from the data of the spin susceptibility,
the systems are described by the free spins induced around 
the impurities in our temperature range.  
Similarity of $C$ to the undoped system implies that the bulk gap 
structure in the excitation spectrum remains unchanged.
At low enough temperatures, 
small interaction between these localized spins is 
expected to yield small peak in the specific heat
to release the entropy, 
which is beyond the temperature range of our QMC simulations.

As is seen in Fig.\ref{fig:sph}, 
at $x>x_{\rm c}$, 
the specific heat at low temperatures ($T\simle 0.1$) is enhanced 
and looks qualitatively different from the case of low doping.
The low temperature tails of the specific heat appears to follow 
$T$-linear dependence, as is pronouncedly visible for 16-impurity doping.
The $T$-linear dependence is 
characteristic to the 1D system with gapless excitations.
In fact, for 1D AFH, from the Bethe Ansatz calculation,  
low temperature specific heat is given as $C=(2/3J)T$ 
\cite{1DAFH_sph} per site.
This value fits well for the $T$-linear part observed at $x>x_{\rm c}$.
It implies that below a crossover 
temperature $T_{\rm cr}$, all the spins become strongly coupled 
similarly to the 1D AFH model.  At $x>x_{\rm c}$, 
$T_{\rm cr}$ increases more or less 
linearly with the doping concentration, namely, 
$T_{\rm cr} \sim 0.04, 0.06$ and 0.09 for $n=8,10$ and $16$, respectively.
%
In fact, when the entropy of $n$ localized spins starts released 
below $T_{\rm cr}$ with $T$-linear specific heat $C\sim(2/3J)T$,
$T_{\rm cr}$ is estimated to be $T_{\rm cr}\sim(3Jx/2)\ln 2$.
This is consistent with the numerical data for $x>x_{\rm c}$.
Although the exponent of the specific heat $\alpha$ defined 
from $C \propto T^{\alpha}$ at asymptotically low temperatures 
may be different from 1 for the randomly distributed impurities, 
the residual entropy in our temperature range may be similar 
between regular and random distribution of impurities.
Therefore, the specific heat 
at $x>x_{c}$ is consistent with the experimental 
observation in ${\rm Zn}$-doped ${\rm SrCu_{2}O_{3}}$.
The peak in the specific heat observed below $T_{\rm cr}$ is 
consistent with temperature dependence of $\chi$, where 
localized spins become strongly coupled below $T_{\rm cr}$.
From the calculated specific heat, 
for ${\rm Zn}$ doped ladder system,
we predict that the specific heat linear in $T$ 
should decrease at $T_{\rm cr}$ when we increase temperature further 
before the exponential growth due to the bulk spin gap appears. 
In the experiments, the linear specific 
heat is observed even at the concentration as low as $x\sim 0.01$
around 10 K.  At least in the temperature of the order of 10K, 
we could not identify the linear specific heat for 
$x<x_{\rm c}$.  This discrepancy may be due to 
the three dimensional coupling.
In fact the AF order appears to be set 
in around the similar temperature.

\begin{figure}
\epsfxsize=8.2cm \epsfbox{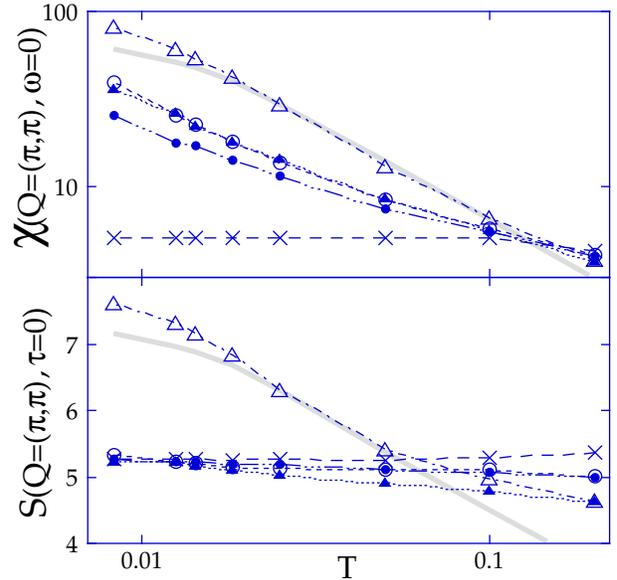}
\caption{ Equal time spin correlation $S$($Q$=($\pi$,$\pi$),$\tau$=0)
and antiferromagnetic suscpetibility $\chi$($Q$=($\pi$,$\pi$),$\omega$=0)
are plotted for 0($\times$), 4($\circ$), 8($\triangle$) 
impurities with regular distributions while bold curves show the case of 
128-site pure AFH chain.
Filled sysmbols show the case of 4(circle) and 8(triangle) 
impurities with random distributions.
}
\label{fig:AFcrl}
\end{figure}

In Fig.\ref{fig:AFcrl}, we plot 
$T$-dependence of equal time AF spin correlation 
$S$($Q$=($\pi$,$\pi$),$\tau$=0) = $N_{s}^{-1}\sum_{i,j} 
e^{-iQ\cdot(r_{i}-r_{j})}\langle S^{z}_{i}S^{z}_{j}\rangle$ 
and AF spin susceptibility
$\chi$($Q$=($\pi$,$\pi$),$\omega$=0) = 
$N_{s}^{-1}\int_{0}^{1/T}d\tau$
$\sum_{i,j} e^{-iQ\cdot(r_{i}-r_{j})} 
\langle S^{z}_{i}(\tau)S^{z}_{j}(0)\rangle$
for both regular and random impurity distributions.

For the regularly distributed case, in contrast to the 4-impurity case,
8-impurity case shows good coincidence with 1D AFH for $T$ $\simle$0.1, 
while $S(Q$=($\pi$,$\pi$),$\tau$=0)
shows similar $T$-dependence as 0-impurity case for 4-impurity case.
It turns out that at $x<x_{\rm c}$, spin gap nature of the ladder system 
is not changed, and the correlation decays exponentially 
in the temperature range studied.
The impurity 
spins become coupled with presumable power-law decay of
correlations only below $T_{\rm cr}$. 
At $x<x_{\rm c}$,
$T_{\rm cr}$ decreases exponentially fast 
with decreasing $x$ and 
it disappears from experimentally accessible temperature 
range quickly because the effective exchange coupling 
between neighboring localized spins confined around the 
impurities should decrease exponentially with the decrease 
in the doping concentration $x$. 
However at $x>x_{\rm c}$, spin correlations 
follow the power law decay,
which again indicates that all the spins become strongly 
coupled as in the 1D AFH chain.  In fact in this regular 
distribution of impurities, spin correlations in real space
have quantitatively similar values to the case of 
1D AFH chain with the same power $\propto 1/r$.
An important point is that in this high-doping region, 
not only the localized spins confined around impurities 
but all the spins in the system take part in this
enhancement of spin correlation.
This is consistent with the results of the magnetic susceptibility and 
the specific heat.


\begin{figure}
\epsfxsize=8.2cm \epsfbox{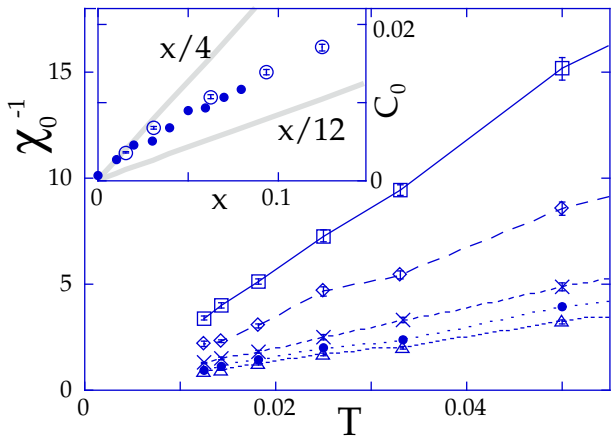}
\caption{ 
$T$-dependence of inverse uniform susceptibility 
for 2($\Box$), 4($\Diamond$), 8($\times$), 12($\bullet$) and 16($\triangle$)
impurities with random distribution on 64$\times$2-site system.
The inset shows the Curie constants obtained from our calculation($\circ$)
and the experiment($\bullet$).
}
\label{fig:curie}
\end{figure}


Next we discuss the case of random impurity distribution.
Residual entropy is expected to be similar to the case of regular 
impurity distribution in our temperature range. 
However, as we see in Fig.\ref{fig:curie}, the uniform susceptibility has 
qualitatively different behavior.  For all impurity concentrations, 
even at $x>x_{c}$, the Curie-like susceptibility is observed. 
In fact, from the arguments by Sigrist and Furusaki,\cite{SigFuru}
at $T\simle \Delta$(spin gap of pure ladder system), 
Curie behavior of uniform susceptibility is 
expected with two different Curie constants $x/4$ and $x/12$ 
with a crossover between these two Curie behaviors around $T$=$T^{*}$.
Here $T^{*}$ is the order of average coupling between impurity spins.
Below $T^{*}$ all impurity spins couple with each other and
the system shows the spin moment due to the fluctuating imbalance 
of the number of impurities on A and B sublattices.
In the relevant temperature range in the experiment $\sim 0.01J(<\Delta)$, 
our calculated Curie constants seem to lie between $x/4$ and $x/12$ 
and is consistent with the experimental observation.\cite{Exprm1}
In terms of the argument by Sigrist and Furusaki,\cite{SigFuru}
it implies that there is a wide intermediate temperature region 
between these two Curie behaviors.


For the case of randomly distributed impurities in Fig.\ref{fig:AFcrl}, 
the growths of $S$($Q$=($\pi$,$\pi$),$\tau$=0) 
and $\chi$($Q$=($\pi$,$\pi$),$\omega$=0) at low temperatures are 
suppressed both from the cases of regularly distributed case 
and the 1D AFH case.
$S$($Q$=($\pi$,$\pi$),$\tau$=0) shows very weak temperature 
dependence in the temperature range studied.
$\chi$($Q$=($\pi$,$\pi$),$\omega$=0) appears to follow approximately 
power-like divergence $\propto T^{-a}$ ($a$ $\sim$ 0.7) with weaker
divergence than the case of 1D AFH ($\propto$ $T^{-1}$).
Although the AF correlation is clearly enhanced from the pure ladder, 
its enhancement is not as large as the 1D AFH case.
This numerical result does not exclude stronger enhancement of
AF correlation at lower temperatures.
However in the experimentally relevant temperature range $T\sim 0.01J$,
this result has shown that the argument of Nagaosa et al.\cite{Nagaosa}
is not applicable because they predict a more singular divergence of 
$\chi$($Q$=($\pi$,$\pi$),$\omega$=0) than 1D AFH.


In summary, effects of nonmagnetic impurities on a spin ladder are 
investigated by the quantum Monte Carlo method with 
loop cluster algorithm. Single impurity causes 
the formation of a spin-1/2 localized spin confined 
around the impurity site.
This impurity spin generates 
a cluster of the size $\xi \sim 3$ around it with 
static AF correlation, which is the 
origin of the enhancement of the AF 
correlation.  From the results from regularly and 
randomly distributed impurities, a rather sharp but 
continuous crossover concentration ($x_{\rm c} \sim 0.04$) 
is identified.  Above $x_{\rm c}$, all the spins are 
strongly coupled with rather uniform and power-law 
decay of the AF correlation.
Below $x_{\rm c}$, the impurity spins are only weakly 
coupled and is expected to 
give rise to power-law decay of the correlations between impurity spins 
only at sufficiently low temperatures. 
Above $x_{\rm c}$ our results are consistent with 
experimental results in terms of two aspects.
One is a large Curie-like susceptibility with the Curie constant $\sim x$.
The other is a small residual entropy and $T$-linear specific heat
with $\gamma$ similar to the per spin value of 1D AFH chain. 
However, it seems to be hard to keep the same behavior 
in the temperature range of $0.01J$ at $x=0.01$.
We speculate that the three dimensional coupling plays 
an important role in keeping the similar behavior even 
at $x$= 0.01.  
The AF correlation is strongly enhanced as compared to the pure ladder.
However, its enhancement is not as large as the 1D AFH chain.

The authors would like to thank H.~Takagi, M.~Sigrist, N.~Furukawa, 
N.~Katoh and Y.~Motome for enlightening discussions.
This work is financially supported by a Grant-in-Aid
for Scientific Research on Priority Area, \backquote Anomalous
Metallic State near the Mott Transition" from the Ministry of
Education, Science and Culture.
The numerical calculations were carried out using facilities of 
Supercomputer Center of the Institute for Solid State Physics, 
Univ. of Tokyo.

\end{document}